\documentclass[10pt,twocolumn,twoside]{IEEEtran}
\usepackage{amsmath,amsfonts}
\usepackage{algorithm}
\usepackage{algpseudocode}
\usepackage{amsmath}     
\usepackage{amssymb}      
\usepackage{array}
\usepackage[caption=false,font=normalsize,labelfont=sf,textfont=sf]{subfig}
\usepackage{textcomp}
\usepackage{stfloats}
\usepackage{url}
\usepackage[utf8]{inputenc}
\usepackage{verbatim}
\usepackage{booktabs}
\usepackage[table]{xcolor}   
\usepackage{graphicx}
\usepackage{cite}
\usepackage{verbatim}  
\usepackage{color} 
\usepackage{amsmath}
\usepackage{float}
\usepackage{array}
\usepackage[utf8]{inputenc}
\begin{document}
%\title{Data-Driven Frequency-Domain Equalization in Molecular Communications}
%\title{Learning to Equalize: LSTM-Based Frequency-Domain Signal Recovery for Molecular Communications}

\title{Learning to Equalize: Data-Driven Frequency-Domain Signal Recovery in\\ Molecular Communications}
\author{\normalsize {Cheng~Xiang,
Yu~Huang,~\IEEEmembership{Member,~IEEE,}
Miaowen~Wen,~\IEEEmembership{Senior Member,~IEEE,}\\
Weiqiang Tan, 
and  Chan-Byoung Chae,~\IEEEmembership{Fellow,~IEEE}} 

\thanks{
This work was supported in part by the National Natural Science Foundation of China under Grants 62571148, 62471183, and 62471152, respectively, in part by GuangDong Basic and Applied Basic Research Foundation under Grant 2025A1515011363, and in part by the Guangzhou Science and Technology Plan Project under Grants 2025A03J3119, 2025A04J7063. The work of C.-B. Chae was supported in part by the Korea government under National Research Foundation of Korea (NRF) Grants RS-2023-00208922 and 2022R1A5A1027646. (Corresponding authors: Yu Huang; Chan-Byoung Chae)

X. Cheng and Y. Huang are with the Research Center of Intelligent Communication Engineering, School of Electronics and Communication Engineering, Guangzhou University, Guangzhou 510006, China (e-mail: 2112330087@e.gzhu.edu.cn; yuhuang@gzhu.edu.cn).\par
M. Wen is with Guangdong Provincial Key Laboratory of Short-Range Wireless Detection and Communication, School of Electronic and Information Engineering, South China University of Technology,
Guangzhou 510641, China (e-mail: eemwwen@scut.edu.cn).\par
W. Tan is with the School of Computer Science and Cyber Engineering, Guangzhou University, Guangzhou 510006, China (e-mail: wqtan@gzhu.edu.cn).\par
C.-B. Chae is with the School of Integrated Technology, Yonsei University, Seoul 03722, South Korea (e-mail: cbchae@yonsei.ac.kr).
}}
%\markboth{Submitted to IEEE Wireless Communications Letters}{Submitted to IEEE Wireless Communications Letters}
\maketitle
\begin{abstract}

In molecular communications (MC), inter-symbol interference (ISI) and noise are key factors that degrade communication reliability. Although time-domain equalization can effectively mitigate these effects, it often entails high computational complexity concerning the channel memory. In contrast, frequency-domain equalization (FDE) offers greater computational efficiency but typically requires prior knowledge of the channel model. To address this limitation, this letter proposes FDE techniques based on long short-term memory (LSTM) neural networks, enabling temporal correlation modeling in MC channels to improve ISI and noise suppression. To eliminate the reliance on prior channel information in conventional FDE methods, a supervised training strategy is employed for channel-adaptive equalization. Simulation results demonstrate that the proposed LSTM-FDE significantly reduces the bit error rate compared to traditional FDE and feedforward neural network–based equalizers. This performance gain is attributed to the LSTM’s temporal modeling capabilities, which enhance noise suppression and accelerate model convergence, while maintaining comparable computational efficiency.

\end{abstract}


\section{Introduction}
\IEEEPARstart{M}{olecular} communication (MC) is an emerging paradigm that enables information transmission at the nanoscale. Unlike traditional wireless systems that rely on electromagnetic waves, MC employs chemical or biological molecules as carriers, offering high biocompatibility and excellent energy efficiency~\cite{farsad2016comprehensive,akyildiz2015internet}. These characteristics make MC highly promising for applications in biomedical engineering, such as targeted drug delivery and in-body monitoring~\cite{chae_iobnt}.

 Practical deployment of MC, however, faces significant challenges. Noise processes in MC differ fundamentally from those in electromagnetic communication. Molecular noise includes both signal-independent and signal-dependent components, the latter arising from fluctuations in molecule concentration and stochastic receptor binding~\cite{5713270}. In addition, signal propagation depends on diffusion and other physical mechanisms (e.g., fluid flow, chemical reactions)~\cite{8742793}, making channels highly sensitive to environmental variations and prone to random delays~\cite{farsad2018molecular}. Moreover, temporal overlap between consecutive symbols leads to severe inter-symbol interference (ISI), degrading detection accuracy. To mitigate ISI, time-domain equalization (TDE) methods, such as MMSE and decision-feedback equalizers, have been extensively studied in MC. While MMSE equalizers reduce symbol errors, their quadratic complexity with respect to ISI length limits real-time applicability~\cite{kilinc2013receiver}. Decision feedback equalizers can further suppress residual interference but suffer from error propagation in low-SNR regimes~\cite{molemimo}. Alternative strategies, including controlled molecule release and MIMO techniques, have also been explored. Nevertheless, these approaches remain constrained by computational overhead and the requirement of accurate channel estimation.

Frequency-domain equalization (FDE) provides an efficient alternative to TDE for ISI mitigation~\cite{huang2021frequency}. Unlike TDE, whose complexity scales with ISI length, FDE leverages frequency-domain processing to reduce computational load while achieving comparable performance, particularly at high data rates. However, like TDE, FDE critically depends on accurate channel state information (CSI). The time-varying nature of MC channels, stochastic perturbations from Brownian motion, and complex noise statistics complicate precise channel estimation, thereby limiting the effectiveness of model-based equalizers.

Data-driven machine learning (ML) approaches have recently emerged as a promising solution. By training on large datasets, ML methods can inherently capture nonlinear and time-varying channel characteristics without explicit analytical modeling. Furthermore, they adapt to variations in noise statistics, improving robustness and ensuring reliable detection across diverse channel conditions. Yet, existing ML-based equalizers in MC have almost exclusively focused on time-domain representations.

In contrast to prior MC studies where ML has been exclusively applied in the time domain, we propose an LSTM-FDE framework that reformulates signal recovery in the frequency domain. This design is motivated by progress in computer vision~\cite{Xu_2020_CVPR}, which demonstrated that frequency-domain features can serve as effective inputs for neural networks and even outperform conventional spatial-domain learning with minimal structural changes. Enlightened by its success and the effectiveness of the FDE paradigm for low-complexity MC signal detection~\cite{huang2021frequency}, we present, to the best of our knowledge, the first frequency-domain learning framework for MC as illustrated in Fig. 1, where the conventional detector is replaced with an LSTM-based network, which models temporal dependencies in molecular signals while reducing the overall computational complexity. Overall, the main contributions of this letter can be summarized as follows:
\begin{itemize}
  \item[$\bullet$] We introduce a data-driven frequency-domain framework for signal recovery in MC, extending ML-based detection beyond conventional time-domain approaches.  
  \item[$\bullet$] A low-complexity LSTM-FDE detector is proposed for MC systems, which reduces computational complexity relative to its time-domain counterparts.
  \item[$\bullet$] We provide a comprehensive evaluation of error performance by benchmarking LSTM-FDE against FNN-FDE and GRU-FDE under various transmission rate scenarios and noise conditions, demonstrating its improved robustness and generalization.  
\end{itemize}

The remainder of this letter is organized as follows. Section~II describes the system model. Section~III introduces the proposed LSTM-FDE and comparative methods. Section IV presents simulation results and performance analysis. Finally, Section V concludes the letter.

\begin{figure}[t]
\centerline{\includegraphics[width=0.50\textwidth]{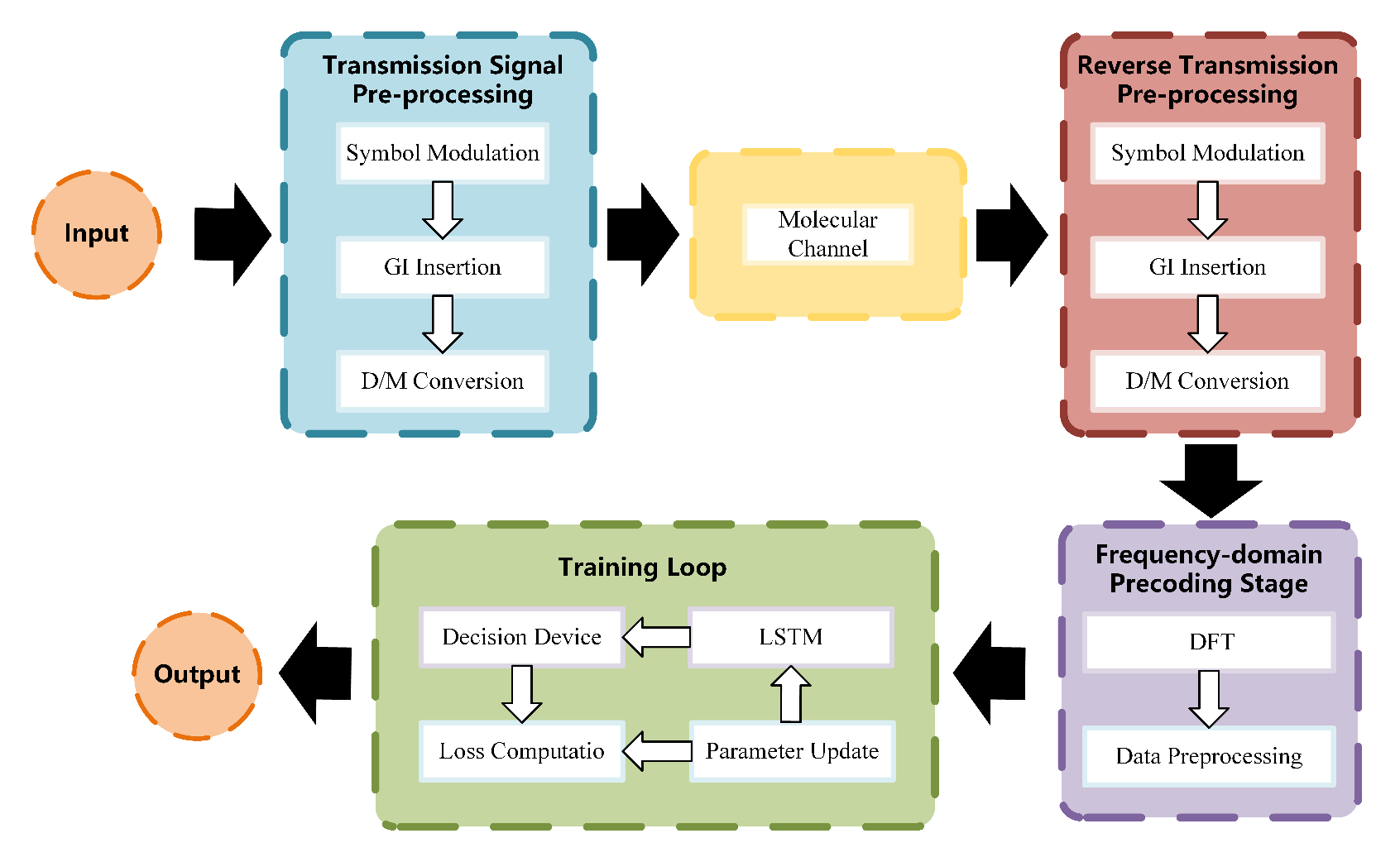}}
\caption{Structure diagram of LSTM-FDE.}
\label{fig1}
\end{figure}

\section{System Model}

This letter employs a classical single-input single-output MC model. The core components consist of a point-source transmitter, a free-diffusion channel, and a passive spherical receiver with radius $r$. The transmitter releases molecules through on-off keying (OOK), with the propagation of information molecules governed by Fick's laws of diffusion in the channel. The receiver passively counts molecules arriving within its surface boundary without absorption or consumption. For the OOK scheme, the transmitted signal at time $t$ is defined as
\begin{equation}
q(t) = \begin{cases}
Q, & \text{bit-}1, \\
0, & \text{bit-}0,
\end{cases}
\end{equation}
where $Q$ denotes the number of molecules released by the transmitter when sending bit “1”, with no molecules emitted for bit “0”.

The channel impulse response (CIR) of MC systems has been rigorously derived in~\cite{8742793}. Here, we assume that molecules are released from the origin and undergo Brownian motion in an unbounded fluid medium, and the receiver is modeled as a sphere of radius \( r \). When the distance $d$ between the transmitter and the receiver satisfies \( d \gg r \), the CIR at time $t$ in an unbounded three-dimensional environment can be derived from Fick’s second law of diffusion~\cite{chae_3d_2014}, and is expressed as
\begin{equation}
h(t)=\frac{V_{\mathrm{r}}}{\left(4\pi Dt\right)^{\frac{3}{2}}}\exp\left(-\frac{d^2}{4Dt}\right),
\end{equation}
where \( D \) represents the diffusion coefficient, and \( V_{\mathrm{r}} = \frac{4}{3}\pi r^3 \) is the volume of the receiver. Considering a symbol duration of \( T_b \) and the ISI length of \( S \), the received signal at time $t$ can be expressed as
\begin{equation}
r(t) = \sum_{s=0}^{S} h(t + s T_b) \cdot q(t - s T_b).
\end{equation}

In our model, two types of noise are considered: signal-dependent noise $n_s(t)$ and signal-independent noise $ n_i(t) $. Signal-dependent noise originates from the inherent randomness of diffusion, which induces fluctuations in the received signal. This component is modeled as a zero-mean Gaussian random variable with signal-dependent variance, i.e., $ n_s(t) \sim \mathcal{N}\left(0, \dfrac{r(t)}{V_r}\right)$. The signal-independent noise, in contrast, reflects the background effect and is modeled as a zero-mean Gaussian random variable with constant variance, i.e., $ n_i(t) \sim \mathcal{N}(0, \sigma^2) $. Accordingly, the received signal is expressed as
\begin{equation}
y(t) = r(t) + n_s(t) + n_i(t).
\end{equation}

Note that the Gaussian model is widely employed in noise modeling for MC, owing to its physical relevance in diffusion-based channels and its analytical tractability~\cite{10904174}.

\section{Data-Driven Methods for Frequency Domain Equalization}

This section introduces three data-driven FDE methods. We first analyze the time complexity of these three methods, and then compare them with other common approaches to assess their performance and applicability.
\subsection{Data-driven Methods}
\subsubsection{FNN Architecture}

A Feedforward neural network (FNN) is a commonly used neural network architecture where each neuron is connected to all neurons in the previous layer. It captures the nonlinear relationships of input signals through a multilayer perceptron (MLP) and learns the mapping between input and target signals for signal recovery. Compared to recurrent neural networks (RNNs), FNNs lack a memory mechanism and struggle with long-term dependencies, but due to their lower computational complexity, they are efficient for short-term dependencies.

\begin{table}[t]
\caption{TIME COMPLEXITY OF DIFFERENT ALGORITHMS} 
\label{table:time_complexity} 
\centering
\begin{tabular}{lc}
\toprule
\textbf{Algorithm} & \textbf{Time Complexity} \\
\midrule
ZF-FDE~\cite{huang2021frequency} & $\mathcal{O}\big(M (\log_2M+1)\big)$ \\
Statistical MMSE-FDE~\cite{huang2021frequency} & $\mathcal{O}\big(M (\log_2M+1)\big)$ \\
$k$-times Iterative MMSE-FDE~\cite{huang2021frequency}  & $\mathcal{O}\big(kM (\log_2M+1)\big)$ \\
FNN-FDE & $\mathcal{O}\big(M\log_2 M + N\nu^2\big)$ \\
LSTM-TDE & $\mathcal{O}\left( M \big( (S+1)\nu + 4N\nu^2\big) \right)$ \\
LSTM-FDE & $\mathcal{O}\big(M(\log_2 M + 4N\nu^2)\big)$ \\
GRU-FDE & $\mathcal{O}\big(M(\log_2 M + 3N\nu^2)\big)$ \\
\bottomrule
\end{tabular}
\end{table}

\subsubsection{GRU Architecture}

The gated recurrent unit (GRU) is a streamlined RNN that merges LSTM’s forget and input gates into a single update gate and replaces the cell state with a hidden state, reducing parameters and training effort. In FDE for MC, the update gate balances residual molecular effects with the current symbol’s response, while the reset gate controls past‑state influence in new activations. This compact architecture captures long‑term dependencies at lower cost than LSTMs, offering an efficient trade‑off between ISI suppression and complexity.

\subsubsection{LSTM Architecture}

LSTM, a variant of RNNs, uses gating mechanisms to capture long‑term dependencies while mitigating gradient issues. In FDE for MC, its gates have clear physical meanings: the forget gate models the diffusion‑dependent decay of residual molecules, the input gate scales the current symbol’s impulse response, and the cell state integrates molecule counts over time. This structure adaptively re‑weights frequency components, improving signal quality in noisy conditions and outperforming feedforward networks in ISI suppression, despite higher training costs.
 
\subsection{Data Pre-processing}

Since FDE operates on discrete-time sequences, the continuous-time received signal in (4) is sampled at physically meaningful instants corresponding to the peak molecular arrival time. Specifically, \( y[n] = y(nT_b + t_p) \), where \( t_p = \frac{d^2}{6D} \)~\cite{huang2021frequency}. It is then transformed into the frequency domain as
\begin{equation}
Y[m] = \sum_{n=0}^{M-1} y[n] \cdot e^{-j\frac{2\pi}{M}nm}, \quad m = 0, 1, \dots, M-1,
\end{equation}
where \( Y[m] \) denotes the complex frequency-domain coefficient at index \( m \), and $M$ refers to the symbol length of each block~\cite{huang2021frequency}. To reduce the amplitude mismatch across samples and stabilize the training of data-driven equalizers, we apply min–max normalization to the magnitude of \( Y[m] \) as
\begin{equation}
\tilde{Y}[m] = \frac{|Y[m]| - min}{max - min + \epsilon},\quad m = 0, 1, \dots, M-1,
\end{equation}
where $max$ and $min$ are the global maximum and minimum magnitudes computed from the entire training set, and \( \epsilon \) is a small constant introduced to prevent division by zero.

The normalized sequence in (6) is the input of the neural networks. The LSTM output is transformed by a fully-connected layer into a scalar value \(o[m]\). For OOK demodulation, a decision module is then applied with a fixed threshold of 0.5 to \(o[m]\), decoding it as ``1" if \(o[m] \geq 0.5\) and as ``0" otherwise.

\subsection{Complexity Comparison}
Table I lists the computational complexity of the equalizers, expressed as the number of real multiplications (or equivalently complex multiplications) required to process a frequency-domain block of $M$ symbols. Here, we consider $N$‑layer neural networks with $\nu$ neurons in each layer for FNN-FDE, $k$ is the iteration count for iterative MMSE‑FDE, and $S$ is the ISI length. 

Based on Table I, model-based FDEs attain optimal efficiency with a complexity of \(\mathcal{O}(M(\log_2M+1))\), primarily driven by FFT operations, but it relies on accurate CSI, whose acquisition is often costly. In contrast, data-driven FDE methods bypass online channel estimation through offline training. Among these, FNN-based FDE exhibits the lowest complexity, approximated as \(\mathcal{O}(M\log_2 M + N\nu^2)\), but suffers from limited ISI suppression capability.

The LSTM-FDE and GRU-FDE architectures, which incorporate gating mechanisms for enhanced channel adaptation, present slightly higher but still competitive complexities, estimated at \(\mathcal{O}(M(\log_2 M + 4N\nu^2))\) and \(\mathcal{O}(M(\log_2 M + 3N\nu^2))\), respectively. In particular, the performance of both architectures is not affected by the ISI length \(S\). On the other hand, the LSTM-TDE incurs a complexity of \(\mathcal{O}(M((S+1)\nu + 4N\nu^2))\), which scales linearly with \(S\) due to its input structure of \((S+1)\). Given that MC channels typically exhibit large \(S\), data-driven FDE methods demonstrate a clear computational advantage over their TDE counterparts, offering a more scalable solution under long channel memory conditions.                                   

\section{Simulation Setup and Results Analysis}

In this section, we present the simulation parameter settings and evaluate the bit error rate (BER) performance (y-axis) across various equalizers, with the number of emitted molecules (x-axis) defined as in (1).

\subsection{Simulation Setup}
Based on the system model, we generate a dataset with 4,000,000 randomly generated binary symbols (“0” or “1”) and perform data pre-processing as described in Section~III.~B. The resulting dataset is divided into training, validation, and test sets with ratios of $60\%$, $20\%$, and $20\%$, respectively. This preprocessed data is used to train an LSTM-based detection system in a supervised manner. Since the input features are in the frequency domain, the LSTM effectively functions as a learnable FDE, whose trainable parameters can be expressed as
\begin{equation}
 \mathbf{W} = \{ \mathbf{W}_f, \mathbf{W}_i, \mathbf{W}_c, \mathbf{W}_o \},
\end{equation}
where \( \mathbf{W}_f \), \( \mathbf{W}_i \), \( \mathbf{W}_c \), and \( \mathbf{W}_o \) denote the input weight matrices associated with the forget gate, input gate, cell-state update, and output gate of the LSTM, respectively. The analysis focuses on this subset of parameters for simplicity, while other components of the full LSTM model are not explicitly included. %The bias terms are denoted by \( \mathbf{b}_* \). The recurrent weights \( \mathbf{U} \) are optimized during training, allowing the network to model temporal dependencies in the frequency-domain sequence through feedback connections.

The neural network is trained without explicit CSI by minimizing a binary cross-entropy loss, which quantifies the error between the predicted bit probability and the true binary label. Based on this per-sample loss, the mini-batch loss function is defined as
\begin{equation}
\mathcal{L} = -\frac{1}{B}\sum_{\iota=1}^{B} \left[ b_\iota \log p_\iota + (1-b_\iota) \log (1-p_\iota) \right],
\end{equation}
where \( B \) denotes the batch size, $b_\iota\in\{0,1\}$ is the ground-truth bit of the $\iota$-th sample, and $p_\iota$ is the predicted probability of the $\iota$-th sample. To minimize \( \mathcal{L} \) with respect to the parameters \( \mathbf{W} \), the gradient \( \nabla_\mathbf{W} \mathcal{L} \) can be computed over the mini-batch using backpropagation through time (BPTT).

\begin{figure}[t]
\centerline{\includegraphics[width=0.45\textwidth]{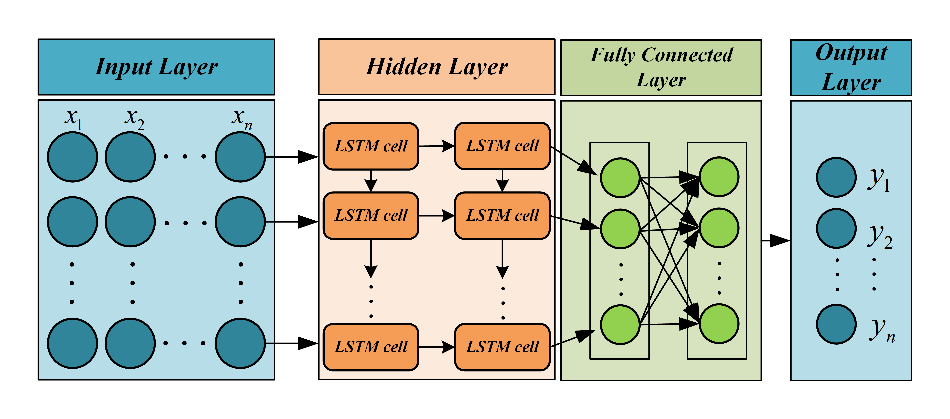}}
\caption{Block diagram of the LSTM-FDE training process.}
\label{Figure.2}
\end{figure}

The parameters are subsequently updated using the Adam optimizer, whose adaptive learning rate is well-suited for the time-varying molecular channel. This joint optimization, driven by BPTT and Adam, enables the equalizer coefficients to converge in a detection-oriented manner, circumventing the need for explicit channel modeling. The overall procedure is shown in Fig.~2{\footnote{The complete pseudo-code and implementation details are available in the extended version on arXiv (arXiv:2509.11327v2 [q-bio.SC]; \url{https://doi.org/10.48550/arXiv.2509.11327}).}}, with FNN and LSTM configurations summarized in Table II. 

\begin{table}[t]
\caption{Comparison of Model Hyperparameter Configurations}
\label{table:model_configs}
\centering
\resizebox{\linewidth}{!}{
\begin{tabular}{lccc}
\toprule
\textbf{Parameter} & \textbf{LSTM Model} & \textbf{GRU Model} & \textbf{FNN Model} \\
\midrule
Optimizer & Adam & Adam & Adam \\
Learning Rate & 0.001 & 0.001 & 0.002 \\
Training Epochs & 30 & 30 & 30 \\
Hidden Units/Layer Structure & 50 & 50 & [256, 128] \\
Activation Function & Tanh & Tanh & ReLU \\
Recurrent Layers & 2 & 2 & -- \\
Batch Size & 64 & 64 & 128 \\
\bottomrule
\end{tabular}
}
\end{table}
%\textit{Proof:} See Appendix.

\subsection{Numerical Simulation Results}

\begin{figure}[t]
\centerline{\includegraphics[width=0.45\textwidth]{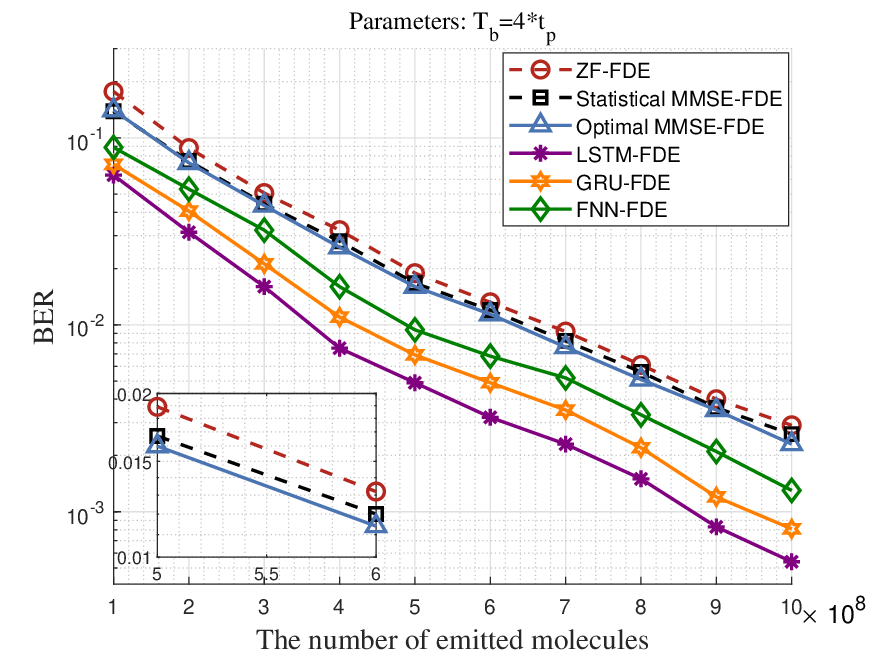}}
\caption{BER performance comparison of various FDEs when the symbol duration $T_b = 4*t_p$.}
\label{Figure.3}
\end{figure}

\begin{figure}[t]
\centerline{\includegraphics[width=0.45\textwidth]{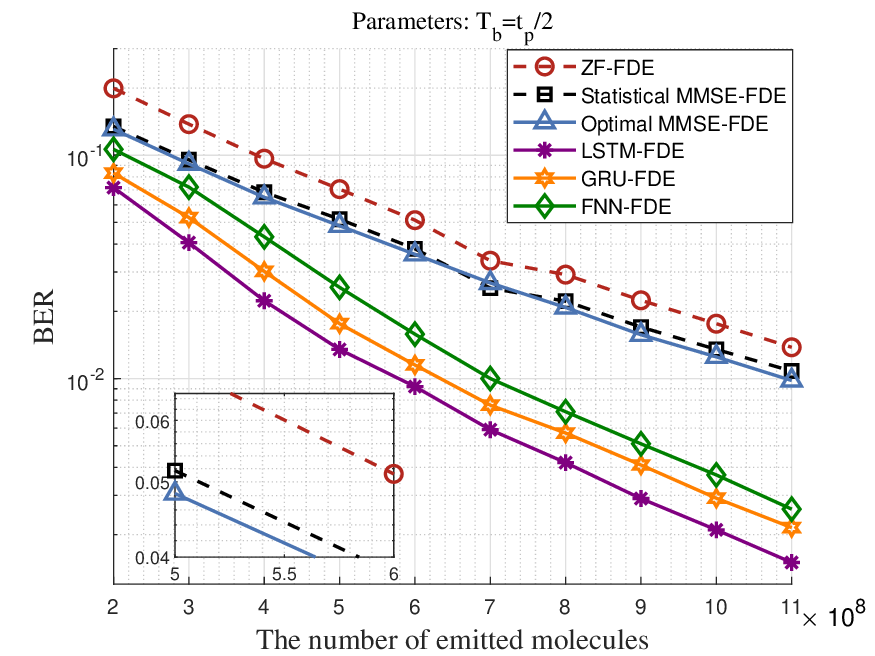}}
\caption{BER performance comparison of various FDEs when the symbol duration \(T_b = t_p/2\).}
\label{Figure.4}
\end{figure}
Figures~3 and 4 compare the BER performance of various equalizers under different symbol durations. Fig.~3 corresponds to a slow transmission case with \(T_b = 4 * t_p\), where model-based FDEs generally achieve lower BER with longer symbol duration. ZF-FDE performs the worst due to the effect of noise amplification, while MMSE-FDE outperforms statistical MMSE-FDE at the cost of higher complexity. Data-driven FDEs consistently surpass model-based approaches. LSTM-FDE achieves the lowest BER, benefiting from its gating mechanism for effective information flow control. GRU-FDE offers intermediate performance, capturing more temporal dependencies than FNN but less long-term modeling capability than LSTM. FNN-FDE exhibits slightly worse performance due to its lack of recurrent structure and constrained temporal feature extraction. 

In contrast, Fig.~4 illustrates the fast transmission case with \(T_b = t_p/2\), where the shortened symbol duration intensifies ISI. Under this condition, ZF-FDE yields the poorest BER among model-based methods, while the optimal MMSE-FDE achieves slightly better performance than its statistical counterpart. Notably, data-driven FDEs are more robust, with LSTM-FDE achieving the highest gain by leveraging temporal dependencies. GRU-FDE and FNN-FDE perform worse, reflecting a trade-off between complexity and accuracy.  
\begin{figure}[t]
\centerline{\includegraphics[width=0.45\textwidth]{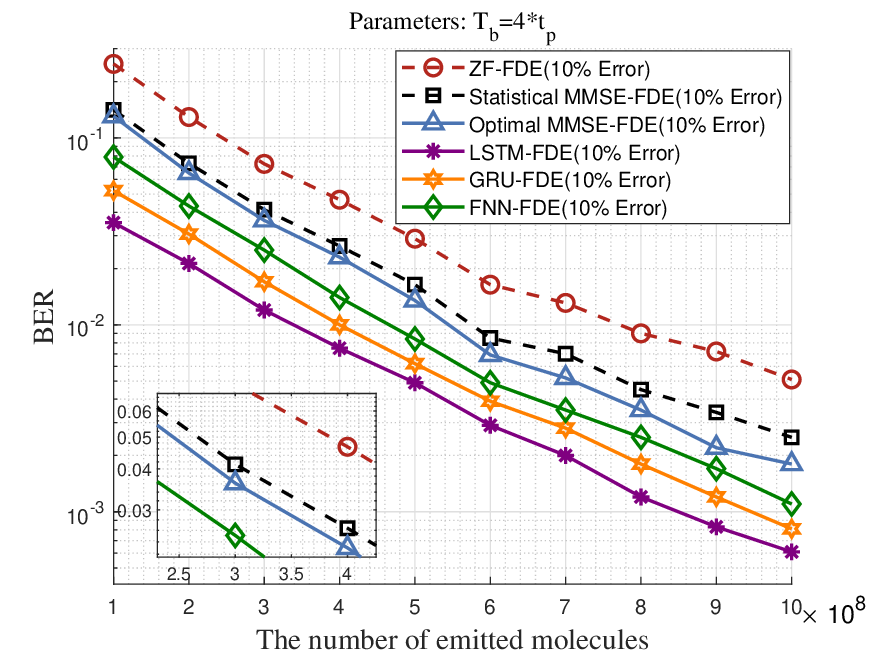}}
\caption{BER performance comparison of various equalizers under channel estimation errors.}
\label{Figure.5}
\end{figure}

Accurate channel estimation is crucial for achieving optimal equalizer performance. 
To further examine the impact of channel estimation errors, Fig.~5 compares the robustness of model-based and data-driven equalizers under CIR amplitude perturbations of $\pm 10\%$. BER decreases as more transmitted molecules, yet the improvement rate varies across schemes. Model-based ZF-FDE and MMSE-FDE, which are highly 
sensitive to estimation errors, demonstrate a less rapid reduction in BER under mismatch. 
In contrast, data-driven approaches show greater robustness, with LSTM-FDE achieving 
the fastest BER decline due to its strong temporal-dependency modeling and noise suppression, 
followed by GRU-FDE. FNN-FDE offers limited gain owing to the lack of memory mechanisms. 
These results highlight that, under channel mismatch, LSTM-FDE can exploit increased molecule 
counts more effectively to enhance reliability.

Figure 6 compares the performance of LSTM‑FDE and LSTM‑TDE at varying symbol rates. In the slow transmission case, LSTM‑TDE performs slightly better than LSTM‑FDE, but the margin is small. On the contrary, in the fast transmission case, LSTM‑FDE’s block‑processing strategy yields a clear advantage, delivering substantially lower BER than LSTM-TDE.  
\section{Conclusion}
In this letter, we proposed LSTM-FDE, a data-driven FDE scheme for MC systems based on LSTM. By leveraging LSTM’s ability to model temporal dependencies, the method achieves high detection accuracy without explicit channel modeling or estimation, learning directly from data. Simulation results show that LSTM-FDE outperforms both model-based FDE and other data-driven baselines across various symbol rates and under channel estimation errors. Compared with its time-domain counterpart, it provides a better trade-off between performance and complexity, making it suitable for fast transmission scenarios. The approach also exhibits strong robustness in dynamic and noisy channels, highlighting its potential for practical biomedical applications such as in-body monitoring. Future work will explore lightweight recurrent architectures, MIMO extensions, and experimental validation on physical MC platforms.

\begin{figure}[t]
\centerline{\includegraphics[width=0.45\textwidth]{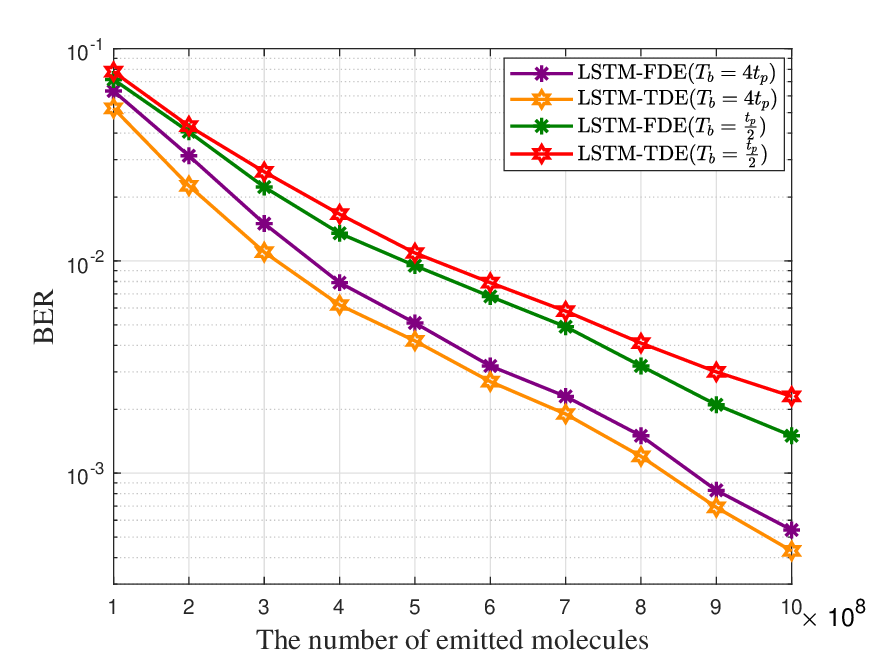}}
\caption{BER performance comparison of LSTM-based TDE and FDE at different symbol durations.}
\label{Figure.6}
\end{figure}
\bibliographystyle{IEEEtran} 
\bibliography{REF}

\newpage
\appendix 

The complete training and inference workflow of the proposed LSTM-FDE is summarized in the following pseudocode.

\section{LSTM-FDE Training and Inference Procedure} 
\begin{algorithm}
\caption{LSTM-FDE Training and Inference Procedure}
\label{alg:lstm_fde_optimized}
\begin{algorithmic}[1]

\Statex Training dataset $\mathcal{D}_{\text{train}} = \{(\widetilde{\mathbf{Y}}_i, b_i)\}_{i=1}^N$
\Statex Hyperparameters (see Table~\ref{table:model_configs})
\Statex Trained LSTM model $\mathbf{W}$, Decoded bits $\hat{b}$

\Procedure{Training}{}
\For{each symbol block in $\mathcal{D}_{\text{train}}$}
    \State $y(t) \gets \text{received signal according to (4)}$ \Comment{Time-domain received signal}
    \State $y[n] \gets y(nT_b + t_p)$ \Comment{Sampling at peak time}
    \State $\mathbf{Y}[m] \gets \text{FFT}(\mathbf{y}[n])$ \Comment{Frequency domain conversion}
    \State $\widetilde{\mathbf{Y}}[m] \gets \frac{|\mathbf{Y}[m]| - \text{min}}{\text{max} - \text{min} + \epsilon}$ \Comment{Normalization according to (6)}
\EndFor
\State Initialize LSTM with random weights $\mathbf{W} = \{\mathbf{W}_f, \mathbf{W}_i, \mathbf{W}_c, \mathbf{W}_o\}$
\For{epoch $= 1$ to $E$}
    \For{each batch $B$}
        \State $\mathbf{p} \gets \text{LSTM}_{\mathbf{W}}(\widetilde{\mathbf{Y}})$ \Comment{Forward pass}
        \State $\mathcal{L} \gets -\frac{1}{B}\sum_{\iota=1}^{B} \left[b_{\iota} \log p_{\iota} + (1-b_{\iota}) \log(1-p_{\iota})\right]$ \Comment{BCE loss according to (8)}
        \State $\nabla_{\mathbf{W}} \mathcal{L} \gets \text{BPTT}$ \Comment{Backpropagation through time}
        \State $\mathbf{W} \gets \mathbf{W} - \eta \cdot \text{Adam}(\nabla_{\mathbf{W}} \mathcal{L})$ \Comment{Parameter update}
    \EndFor
\EndFor
\State \Return trained model $\mathbf{W}$
\EndProcedure

\Procedure{Inference}{$\mathbf{y}_{\text{test}}[n]$}
\State Preprocess $\mathbf{y}_{\text{test}}[n]$ as in Training \Comment{FFT + normalization}
\State $\mathbf{o} \gets \text{LSTM}_{\mathbf{W}}(\widetilde{\mathbf{Y}}_{\text{test}})$ \Comment{Forward pass}
\State $\hat{b}[m] \gets \mathbf{1}_{\{o[m] \geq 0.5\}}$ \Comment{Threshold decision at 0.5}
\State \Return $\hat{b}$
\EndProcedure
\end{algorithmic}
\end{algorithm}

\end{document}